# Thickness dependence of the interfacial Dzyaloshinskii-Moriya interaction in inversion symmetry broken systems

Jaehun Cho[1], Nam-Hui Kim[1], Sukmock Lee[1], June-Seo Kim[2], Reinoud Lavrijsen[2], Aurelie Solignac[2], Yuxiang Yin[2], Dong-Soo Han[2], Niels J. J. van Hoof[1,2], Henk J. M. Swagten[2], Bert Koopmans[2], and Chun-Yeol You[1]

[1]Department of Physics, Inha University, Incheon 402-751, Republic of Korea

[2]Department of Applied Physics, Center for NanoMaterials, Eindhoven University of Technology, PO Box 513, 5600 MB Eindhoven, The Netherlands

Correspondence and requests for materials should be addressed to C.-Y.Y (email: cyyou@inha.ac.kr) and J.-S.K (email: spin2mtj@gmail.com)

In magnetic multilayer systems, a large spin-orbit coupling at the interface between heavy metals and ferromagnets can lead to intriguing phenomena such as the perpendicular magnetic anisotropy, the spin Hall effect, the Rashba effect, and especially the interfacial Dzyaloshinskii-Moriya (IDM) interaction. This interfacial nature of IDM interaction has been recently revisited because of its scientific and technological potential. Here, we demonstrate an experimental technique to straightforwardly observe the IDM interaction, namely Brillouin light scattering. The non-reciprocal spin wave dispersions, systematically measured by Brillouin light scattering, allow not only the determination of the IDM energy densities beyond the regime of perpendicular magnetization but also the revelation of the inverse proportionality with the thickness of the magnetic layer, which is a clear signature of the interfacial nature. All together, our experimental and theoretical approaches involving double time Green's function methods open up possibilities for exploring magnetic hybrid structures for engineering the IDM interaction.

In the presence of spin-orbit coupling at interfaces on low dimensional magnetic hetero-junction structures, the effect of structural inversion asymmetry leads to an additional anisotropic exchange term, namely the interfacial Dzyaloshinskii-Moriya (IDM) interaction[1-4] (already predicted by Fert in 1980), which is a branch of the Dzyaloshinskii-Moriya (DM) interaction[5,6]. This interfacial phenomenon has been recently re-illuminated and experimentally demonstrated because of its massive potentials to explore new magnetic behaviours such as chiral domain wall (DW) dynamics[7-13] and skyrmions[14-16]. To develop this field of DW devices and skyrmionics (the latter with great promises for superior nanoelectronics devices), experimental tools to extract the magnitude and sign of IDM interaction are urgently required. However, contrary to bulk-type DM interaction measurements[17], recent extensive experiments clearly observed the existence of the IDM interaction, but magnetic field and electric current driven DW dynamics measurements were definitely linked to the perpendicular magnetic anisotropy (PMA)[8-13]. At present, in order to further explore independent by the underlying physics of the IDM interaction without any other linked material parameters, a radically different experimental approach is strongly required.

In this article, we, measure the ferromagnetic layer thickness dependent of IDM interaction *quantitatively* and *qualitatively*. Inelastic light scattering, so-called Brillouin light scattering (BLS), is performed to observe non-reciprocal spin wave (SW) dispersion relations affected by the IDM interaction[18]. The detailed explanation about the BLS is shown in Supplementary Note 1. The advantages of BLS to determine the IDM energy density is described in Supplementary Note 2. Our main findings are twofold: first, the inverse proportionality of the IDM energy densities to the ferromagnetic layer thickness shows that the IDM interaction is purely originated from the interfaces, and second, we present a state-of-the-art quantum-mechanical approach to confirm the asymmetric dispersion relations and the inverse proportionality of the IDM interaction. As representative hetero-structures, Pt/Co/AlO$_x$ and Pt/CoFeB/AlO$_x$ are chosen because these multilayer structures are already predicted to have a large IDM interaction[12].

**Results**

**SW Frequency differences due to the IDM interaction.** Propagating SWs on a magnetic thin film can be localized at the top and bottom surfaces of the ferromagnetic layer when the wavevector $k$ of the SW is perpendicular to the magnetization of the system. This SW mode is namely Damon-Eshbach (DE) mode (often called surface mode)[19] and it is indeed one of the appropriate physical quantities to investigate the interface effect, especially affected by the IDM interaction. In order to realize the DE geometry, we first apply an external magnetic field along the in-plane as depicted in Fig. 1a. Simply, BLS measures the scattered light from two interfaces, which contains photons at frequencies shifted by the frequencies of excited SWs. In this inelastic process, the photon loses its kinetic energy (Stokes process) to create one of quasi-particles (SW in our study) or gains energy (anti-Stokes process) by absorbing one. Consequentially, these spectral components can determine the frequencies and intensities of SWs existing at the point in the sample where the incident light is focused (See Methods).

Usually, the SW frequencies of Stokes and anti-Stokes peaks should be at the same position or slightly different due to the PMA energy difference between top and bottom interfaces of the ferromagnet[20]. However, recent theoretical and numerical calculation proposes a prominent clue that the frequencies and the attenuation lengths of the travelling SWs with opposite wavevectors ($\pm k$) are significantly different due to the IDM interaction, and then these characteristics of the SWs are measurable[21]. For BLS, the frequency difference ($\Delta f$) indicates the mismatch between the frequencies of Stokes and anti-Stokes peaks. We report that a large frequency difference ($\Delta f$ = 1.99 GHz) for Pt/Co(1.2 nm)/AlO$_x$ is obviously observed as shown in Fig. 1b. Now, one crucial fact can be emphasized that the GHz range of the $\Delta f$ is a clear signature of the IDM interaction. The details will be further discussed later.

**Magnetic field dependence.** In order to precisely and systematically investigate this interface effect by means of BLS, two different methods (magnetic field dependence and incident angle dependence) as a function of the thickness of the ferromagnet ($t_{FM}$) are proposed in this work. We now discuss the details of two different methods successively. The DE SW frequencies (dispersions) including the IDM interaction are given as[21]:

$$f_{DE} = f_0\left(M_S, H_{ext}, K_U, A_{ex}, k_x \hat{\mathbf{x}}\right) + p\frac{\gamma D}{\pi M_S}k_x, \tag{1}$$

where, $f_0$ is the SW frequency without the IDM contribution, $H_{ext}$, $K_u$, $A_{ex}$, $\gamma$, $p$ and $k_x = \dfrac{4\pi}{\lambda}\sin\theta$ are the external magnetic field, the magnetic anisotropy, the exchange stiffness, the gyromagnetic ratio, the polarity of the magnetic field ($p = \pm 1$) and the wavevector of the SW, where $\theta$ is the incident angle of the light, respectively. Therefore, the definition of the frequency difference is simply derived as $\Delta f = \left| f_{DE}\left(+k_x\hat{\mathbf{x}}\right) - f_{DE}\left(-k_x\hat{\mathbf{x}}\right) \right| = \frac{2\gamma D}{\pi M_S} k_x$, where $M_s$ and $D$ are the saturation magnetization and the IDM energy density, respectively. This SW dispersion apparently shows that the $\Delta f$ is invariant while the magnetic field increases (or decreases). So, the field dependent measurement allows us to minimize the uncertainties of the measured $\Delta f$. The measured SW frequencies of Stokes and anti-Stokes peaks as a function of the magnetic field for Pt/Co(1.2 nm)/AlO$_x$ are shown in Fig. 2a. Two SW frequencies increase with increasing applied magnetic field, but the $\Delta f$ (the averaged $\Delta f$ is 2.18 GHz) is indeed a constant (see the inset in Fig. 2a). From these field dependent measurements, we can convert to the IDM energy densities because the SW wavevector is fixed at $k_x = 0.0167$ nm$^{-1}$ ($\theta = 45°$). It must be noted that the minimum applied in-plane field is 0.5 T is large enough to pull the magnetization in the plane. Therefore, the observed $\Delta f$ is only for the in-plane magnetization, and we cannot conclude that $\Delta f$ will vanish or not when the magnetization is out-of-plane. Due to the limitation of BLS measurement setup, it is hard to determine $\Delta f$ for the out-of-plane magnetization (see additional Supplementary Note 3).

In many magnetic systems, interface effects can be identified by an inverse proportionality to the ferromagnetic layer thickness such PMA[22], exchange bias[23], switching current density of spin transfer torque[24], the effective field of the interlayer exchange coupling[25], and so on. In this point of view, we systematically measure $\Delta f$ as a function of the thickness of the ferromagnets (Co and CoFeB) and nine data points with different magnetic fields are averaged for each thickness. As shown in Fig. 2b, $\Delta f$ approaches to the origin when $t_{FM}^{-1} \to 0$ for both Co and CoFeB samples by which we consequently confirm that the IDM interaction for our asymmetric structures is purely originated from the interface. For the thinner CoFeB cases ($t_{CoFeB} < 1.6$ nm), the frequency differences deviate from the inverse proportionality. It implies that the non-linear behaviour in Fig. 2b is due to the degradation of the interface quality (see Supplementary Note 4).

**SW propagation direction dependence of the $\Delta f$.** Continuously, we now demonstrate another clear proof that the asymmetric frequency differences are indeed emerged from the interface. In recent previous theoretical work, Cortés-Ortuño[26] claims that the frequency differences $\Delta f$ by the DM interaction can be expressed as:

$$\Delta f(\alpha) \sim \Delta f_0 \sin\alpha, \tag{2}$$

where $\alpha$ indicates the angle between $\mathbf{k}_\parallel$ and $\mathbf{M}$, and $\Delta f_0$ is a frequency difference at $\alpha = \pi/2$. The physical interpretation of Eq. (2) is that the frequency differences $\Delta f$ is created by the energy differences of two propagating SWs for both interfaces. Since the IDM interaction introduces these energy differences, this equation is another clear evidence of the DM interaction, especially for the case of the interface effect. Figure 3 shows the angular dependence of the frequency differences between the angle of SW *k*-vector and the direction of $\mathbf{M}$. Fig. 3a indicates the case of $\alpha = \pm \pi/2$ (usual BLS measurement geometry, i.e. $\mathbf{k}_\parallel \perp \mathbf{H}$) and $\alpha = 0^\circ$ (90° rotation from usual BLS measurement geometry, i.e. $\mathbf{k}_\parallel // \mathbf{H}$). It is clearly shown that $\Delta f$ (=+1.71, -1.73 GHz) are finite and opposite sign for $\alpha = \pm \pi/2$, while $\Delta f$ = 0.11 GHz for $\alpha = 0^\circ$ is less than BLS limitation (~ 0.29 GHz, see the Supplementary Note 5). The systematic angular dependent measurements are shown in Fig. 3b and we overlap the sinusoidal curve from Eq. (2). As expected, they are in good agreement with each other.

**SW *k*-vector dependence.** Furthermore, we measure the dispersion relation of SW (frequency versus wavevector) by varying the incident angle $\theta$ of the probing light which determines the selected SW's wavevector $k_x \hat{\mathbf{x}}$. We note that the magnetic field- and $k_x$-dependent measurements span two different branches of Eq. (1), and those two independent measurements can provide more reliable results in the present study. The dependence of $f_{DE}$ on $k_x$ are plotted in Fig. 4a for various Co thicknesses. The solid lines correspond to linear fit to the experimental results. For all Co thicknesses, the $f_{DE}$ linearly decreases with increasing $k_x$. Following the Eq. (1), $f_0$ and $\Delta f$ varies quadratically and linearly with $k_x$, respectively. However, for the limited range of $k_x$ (-0.03 nm$^{-1}$ < $k_x$ < +0.03 nm$^{-1}$), the $f_0$ are almost constant, accordingly, such significant variation in $f_{DE}$ results from those in $\Delta f$. Therefore, these asymmetric and linear dispersion relations can be regarded as the direct evidence that the $\Delta f$ in our measurement is a consequence of IDM interaction. Recently, the asymmetric SW dispersion relation has been experimentally observed by using spin-polarized electron energy loss spectroscopy in double layer Fe films[27]. For comparison, we also examined the dispersion relation for Pt(4 nm)/Co(0.6 nm)/Pt(4 nm) representing a symmetric structure, where the IDM interaction at the bottom and top interfaces of the FM are known to be approximately of the same magnitude but with the opposite sign, thus leading to zero IDM interaction. Interestingly, for the symmetric structure, no significant IDM interaction is observed (see Supplementary Note 5). Figure 4b shows the $\Delta f$ versus $|k_x|$ for selected Co thickness, $t_{Co}$ = 1.1, 1.2, 1.5, and 1.6 nm. For each film thickness, one obtains a clear linear dependence. From the slopes, we can extract IDM energy density using the relation of Eq. (1) together with the gyromagnetic ratio $\gamma$ and the saturation magnetization $M_s$ deduced from the further BLS measurements.

**The IDM energy densities.** Next, we convert the measured $\Delta f$ to the IDM energy densities for our asymmetric structures as shown in Fig. 5. For Co samples, the measured IDM energy densities ($D_H$ and $D_k$ indicate the IDM energy densities from the field dependence and SW wavevector dependence, respectively.) are in excellent agreement each other (see Fig. 5a). Figure 5b shows the measured IDM energy densities for CoFeB sample. In this case, the effective magnetic anisotropy for all thicknesses is in-plane. The maximum IDM energy density is obtained about $D$ = 1.2 mJm$^{-2}$ ($D$ = 0.7 mJm$^{-2}$) for Pt/Co(1 nm)/AlO$_x$ (Pt/CoFeB(1.6 nm)/AlO$_x$). Recall that the saturation magnetization ($M_s$) is one unique material parameter to convert the IDM energy density in Eq. (1). The saturation magnetization $M_s$ (=1100 kAm$^{-1}$ for Co and 948 kAm$^{-1}$ for CoFeB) are determined by BLS measurement as well (see Supplementary Note 6).

**Numerically calculated SW dispersion relation.** Two types of BLS measurements (magnetic field dependence and $k_x$-vector dependence) are presented so far. From these measurements, we found the inverse proportionality of $\Delta f$, which is a typical signature of the interfacial nature and the asymmetric dispersion relation. One of the main goals of the present work is to demonstrate the SW dispersion relation affected by the IDM interaction and the inverse proportionality not only by experiment but also theoretically. In previous work, theoretical evidences based on atomic-scale models[28,29,30] have been reported. Udvardi et al.[28] predict reciprocal SW dispersion relations for the specific crystallographic orientation in the Fe/W(110) by using first principle calculations, without dipole-dipole interaction and external field, and Costa et al.[26] provide dynamic susceptibilities (SW frequencies, life times, amplitudes) for ± SW vectors in the one or two monolayer (ML) of Fe on the W(110) based on multiband Hubbard model. Cortés-Ortuño and Landeros demonstrate reciprocal SW dispersion relations for different crystallographic classes. Here, we introduce the numerical calculations for asymmetric SW dispersion relations and inverse proportionality by means of the double time Green's function technique, it is useful to study the thickness dependent SW dispersion relations. This technique is well developed in statistical physics[30] and magnetism[31,32]. The Hamiltonian with the IDM interaction for the finite thickness ferromagnetic layer in terms of the spin operators is given by[32,33]:

$$H = -g\mu_B H_{ext} \sum_i S_i^z - \sum_{\langle i,j \rangle} J_{ij} \mathbf{S_i} \cdot \mathbf{S_j} - K_u \sum_{\langle i,j \rangle} S_i^z S_j^z - K_s \sum_{\langle i,j \rangle'} S_i^z S_j^z + \sum_{\langle i,j \rangle''} D_{ij} \hat{\mathbf{z}} \cdot (\mathbf{S_i} \times \mathbf{S_j}), \qquad (3)$$

where, $g$ is the Lande $g$-factor, $J_{ij}$ and $D_{ij}$ are the isotropic inter-atomic Heisenberg and anisotropic DM exchange energies between the $i$-th and $j$-th spins, and $K_u$ and $K_s$ are the bulk and surface uniaxial anisotropy energies. In this model, we use different definition of coordinate system, we set the film normal along the $z$-axis. $\langle i,j \rangle, \langle i,j \rangle'$, and $\langle i,j \rangle''$ denote the summations of the nearest neighbours. Here, $\langle i,j \rangle$ is summation of all spins, $\langle i,j \rangle'$ is for top and bottom interfaces, and $\langle i,j \rangle''$ is only at the bottom interface where we assumed a heavy metal is placed. Therefore, we assume that the DM interaction exists only at the bottom interface. The SW dispersion relations can be obtained by solving Eq. (3). The detailed explanations and full derivations are shown in Supplementary Note 7.

Figure 6a shows numerically calculated SW dispersion relations for a ferromagnetic ML with the IDM interaction term $\delta_0 = SD/J_{ex}a$, where $S$, $a$, and $J_{ex}$ are spin number 1/2, the lattice constant and the exchange energy, respectively. A parabolic SW dispersion relation (black line) is obtained when $\delta_0 = 0$, when $\delta_0$ is non-zero, parabolic SW dispersion relations are shifted as given by Eq. (1) and shown by the red and blue lines for different strengths of $\delta_0$ ferromagnetic ML. As mentioned above, the SW $k_x\hat{\mathbf{x}}$-vector of our BLS setup is limited from 0.0099 nm$^{-1}$ to 0.0205 nm$^{-1}$; the small range is indicated by the green box in Fig. 6a. That is the reason that we obtained only linear behaviours of $f_{DE}$ in Fig. 4a and one can be pointed out that this numerical result can sufficiently support our experimental data. Finally, the inverse proportionality of the IDM energy density as a function of the thickness of the ferromagnetic layer ($t_{FM}$) is shown in Fig. 6b and the inset indicates asymmetric SW dispersion relations for $\delta_0 = 0.1$. In this calculation, we consider the thickness of the ferromagnetic layer from 2 ML to 20 ML. These full numerical SW dispersion relations reflect our experimental observations very well. First, the SW frequencies at $k_x = 0$ increase with increasing $t_{FM}$. The experimental results show the same trend in Fig. 4a. Since the SW frequency is related with the interface PMA energy, it must be increased with increasing $t_{FM}$ (see Supplementary Fig. 5a). Second, the parabolic SW dispersion relations have an additional linear $k_x$. Because the coefficient of a linear $k_x$ term is proportional to the $D_k$, we can directly extract $D_k$ from the SW dispersion relations. Very recently, there is another numerical and theoretical approach about the interface exchange boundary conditions for the classical linear dynamics of magnetization[34]. This profound and accurate prediction also shows the inverse proportionality of the frequency difference and the results are consistent with our experimental and numerical data.

**Discussion**

In conclusion, using a versatile light scattering technique, we have observed the IDM interaction in the inversion symmetry broken systems. The quantitative magnetic layer thickness dependent measurements and careful analysis show the inverse proportionality of the frequency differences and confirm that the IDM interaction is a pure interfacial effect with maximum energy density of 1.2 mJm$^{-2}$ for Co with Pt underlayer. Furthermore, two different measurement methods, the magnetic field dependence and SW wavevector dependence, allow us to obtain identical results. These findings take us a step closer to boosting the IDM interaction leading to (meta-) stable skyrmion state for future data and memory devices. Finally, our brand new numerical calculations confirm the asymmetric SW dispersion relations due to the IDM interaction and the inverse proportionality.

**Methods**

**Thin film deposition.** The sample of Pt(4 nm)/Co(0-2 nm)/AlO$_x$(2 nm) and Pt(4 nm)/Co$_{48}$Fe$_{32}$B$_{20}$(0-2 nm)/AlO$_x$(2 nm) were prepared on Si/SiO$_2$ substrates using DC magnetron sputtering with a base pressure of ~7×10$^{-8}$ mbar. To investigate the thickness dependence of IDM interaction, the ferromagnetic layers were grown in a wedge shape over 2cm wide wafers with the help of an *in-situ* moving shadow mask. AlO$_x$ layer was obtained from plasma oxidation of 2-nm-thick Al layer as deposited on top of the ferromagnetic layers. The plasma oxidation process was carried out for 10minutes in an *in-situ* isolated chamber with a 0.1mbar background pressure of oxygen and a power of 15W.

**Brillouin light scattering.** The samples are pasted on an angle controlled sample holder for the BLS measurement. The BLS spectra are measured by using a (3+3) pass tandem Fabry-Perot interferometer and a *p*-polarized (300 mW power and 532 nm wavelength) single longitudinal mode LASER is used as a light source. The DC external magnetic field is applied parallel to the film surface and perpendicular to the scattering plane. The back-scattered light from the sample is focused and collected. The *s*-polarized light is passed through the interferometer and the photomultiplier tubes[35]. All measurements are performed at room temperature. We use the applied magnetic field (0.01 T ~ 1.18 T) and incident angle of light (25° ~ 60°) corresponding to $k_x$ = 0.0099 ~ 0.0205 nm$^{-1}$ for magnetic field dependence and dispersion relation measurements, respectively. The accumulation time for each spectrum was about 60 min.

**Acknowledgements**

This work is supported by the research program of the Foundation for Fundamental Research on Matter (FOM), which is part of the Netherlands Organisation for Scientific Research (NWO), and National Research Foundation of Korea (Grant No. 2013R1A1A2011936) and KIST institutional program (2E24882).

**Author Contributions** C.-Y.Y. and J.-S.K. conceived the project; sample fabrication by J.-S.K. and Y.Y.; the measurements were performed by J.C., N.-H.K., N.J.J.v.N. and A.S.; data analysis and manuscript preparation were done by J.-S.K., J.C., R.L., D.-S.H., Y.Y. and C.-Y.Y.; numeral calculation were done by C.-Y.Y; all authors discussed the results.

**Additional Information**

**Competing Financial Interests** The authors declare no competing financial interests.

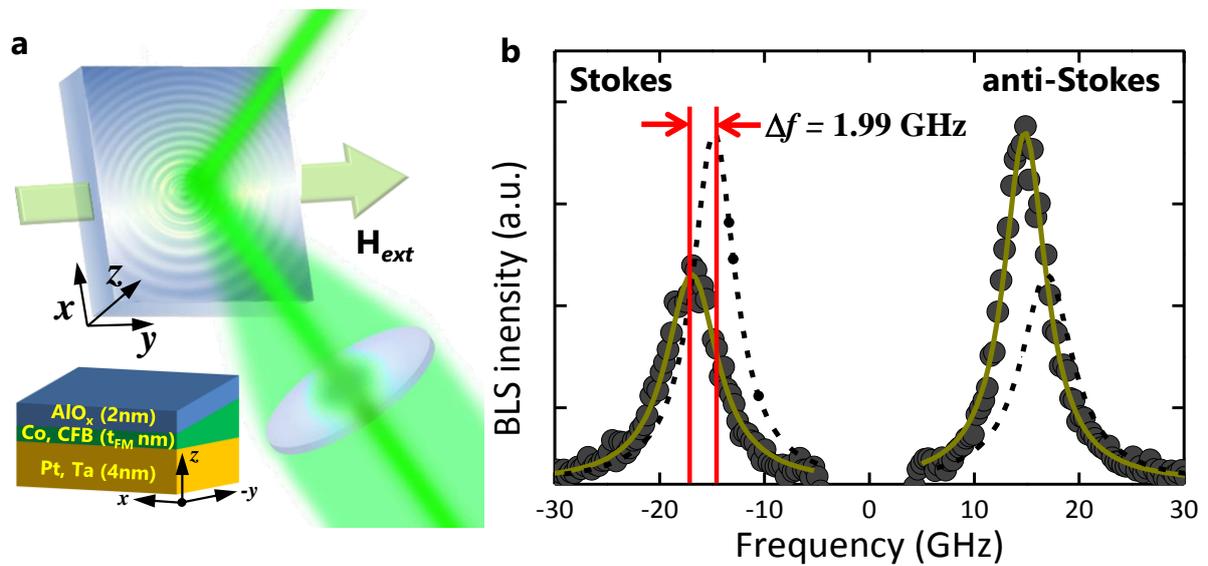

**Figure 1 | The BLS measurements. a**, Schematic configuration of the BLS measurement. The external magnetic field is applied along the *y*-direction and a *p*-polarized laser beam excites two interfaces SWs along the *x*-direction. Inset: schematic picture of wedge-type sample geometry. **b**, The BLS spectrum with a magnetic field $H_{ext}$ = 0.69 T. The incident angle is fixed at $\theta$ = 45° ($k_x$ = 0.0167 nm$^{-1}$). In order to identify the frequency differences ($\Delta f$) between Stokes (negative frequency region) and anti-Stokes (positive frequency region), mirrored curves are drawn as black dashed line. The red vertical lines indicate the centre of the SW frequency and red arrows indicate the $\Delta f$, here 1.99 GHz. The black circles refer to the experimental result and dark yellow solid line is the Lorentzian fitting curve. The data accumulation time for each spectrum is about 60 minutes.

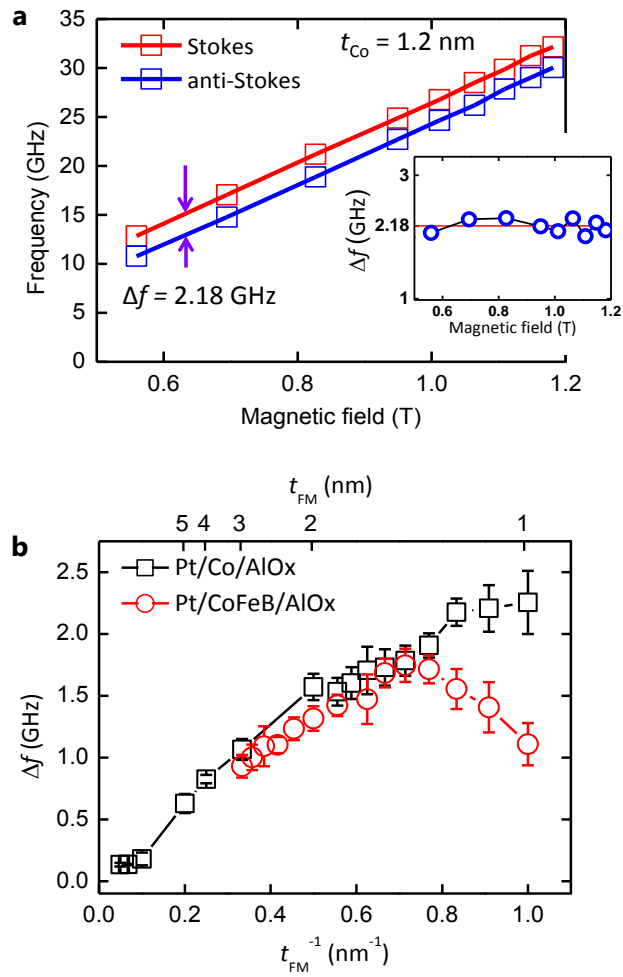

**Figure 2 | The $H_{ext}$ dependence measurement and the $\Delta f$ between Stokes and anti-Stokes peak. a,** Magnetic field dependent BLS measurements at $t_{Co}$ = 1.2 nm. The in-plane magnetic field varies from 0.5 T to 1.2 T and the angle of the incident light is fixed at $\theta$ = 45°. The violet arrows are average $\Delta f$ is 2.18 GHz between Stokes (red squares and line) and anti-Stokes (blue squares and line) peaks. inset: The frequency differences ($\Delta f$) as a function of applied magnetic field. **b,** $\Delta f$ as a function of $t_{FM}^{-1}$ for two different magnetic materials (Co and CoFeB). Black squares and red circles indicate $\Delta f$ for Co and CoFeB, respectively. For these measurements, the incident angle is fixed at $\theta$ = 45°, which corresponds to the $k_x$ = 0.0167 nm$^{-1}$. Error bars correspond to the standard deviations of the BLS measurements.

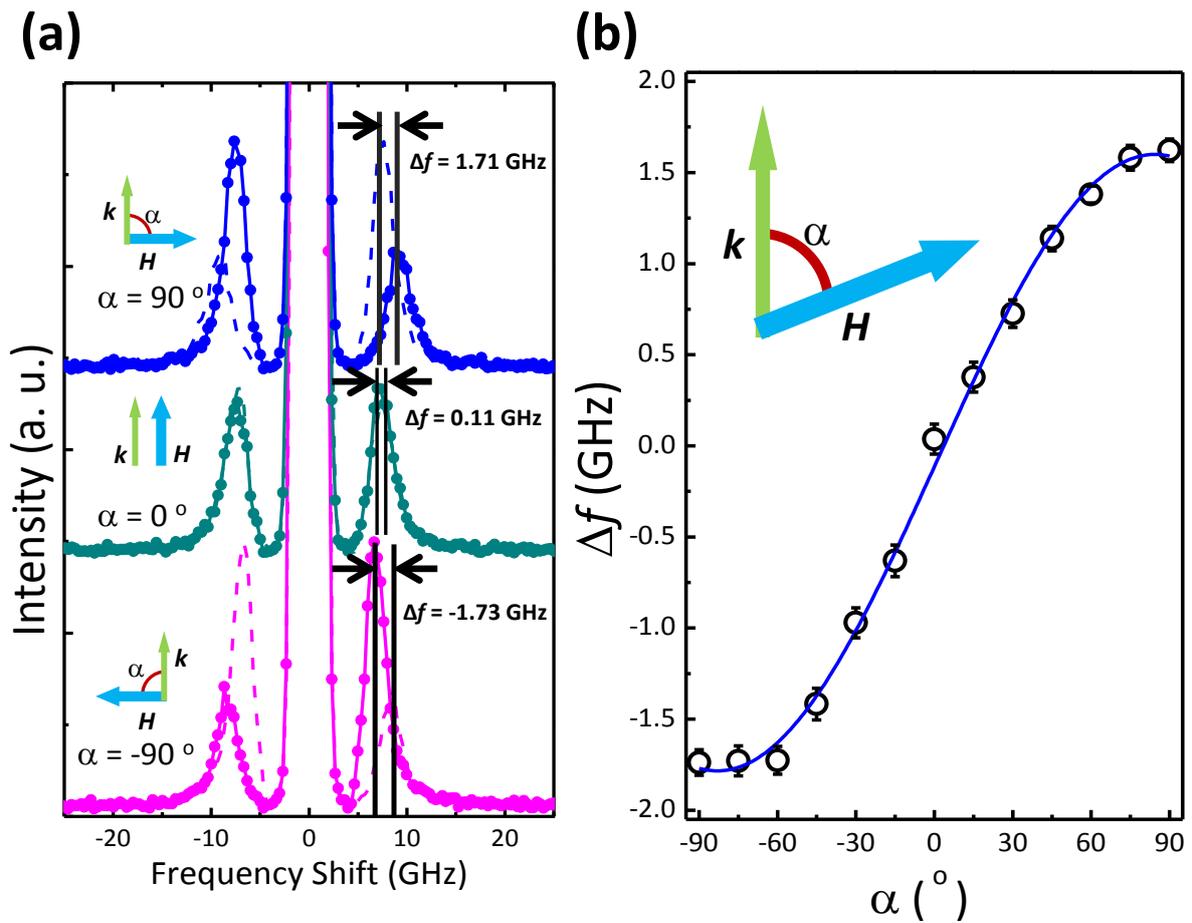

**Figure 3 | The BLS spectra for $\alpha = 90°$ and $0°$, and $\alpha$ dependence of $\Delta f$ for 2.0-nm thick Co. a**, The frequency difference between Stokes and anti-Stokes are clearly observed ($\Delta f$ =+1.71, -1.73 GHz) for $\alpha = \pm 90°$ ($\mathbf{k}_{||} \perp \mathbf{H}$), while the $\Delta f$ (=0.11 GHz) for $\alpha = 0°$ ($\mathbf{k}_{||}$ // $\mathbf{H}$) is less than BLS resolution. The black vertical lines and the black arrows indicate the $\Delta f$ between Stokes and anti-Stokes peaks. The green and blue arrows indicate the directions of the SW wavevector and the applied magnetic field, respectively. **b**, The measured $\Delta f$ as a function of the $\alpha$. The solid line is the fitting curve from Eq. (2). Error bars correspond to the standard deviations of the BLS measurements.

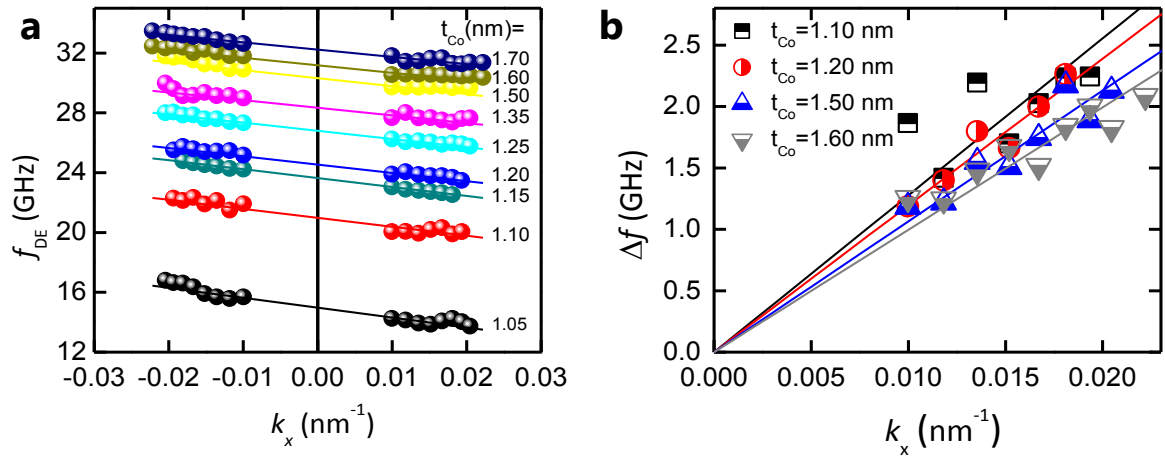

**Figure 4 | The SW dispersion relation and the linearity of the Δ$f$ in each thickness. a,** The asymmetric dispersion relation measured by the BLS for various Co thicknesses. For these measurements, the applied magnetic field is fixed at $H_{ext}$ = 0.915 T. The solid lines correspond to linear fit to the experimental results. For all Co thicknesses, the $f_{DE}$ linearly decreases with increasing $k_x$. These asymmetric and linear dispersion relations can be regarded as the direct evidence that the Δ$f$ in our measurement is a consequence of IDM interaction. **b,** All Δ$f$ and linear fitting lines are visualized in one graph.

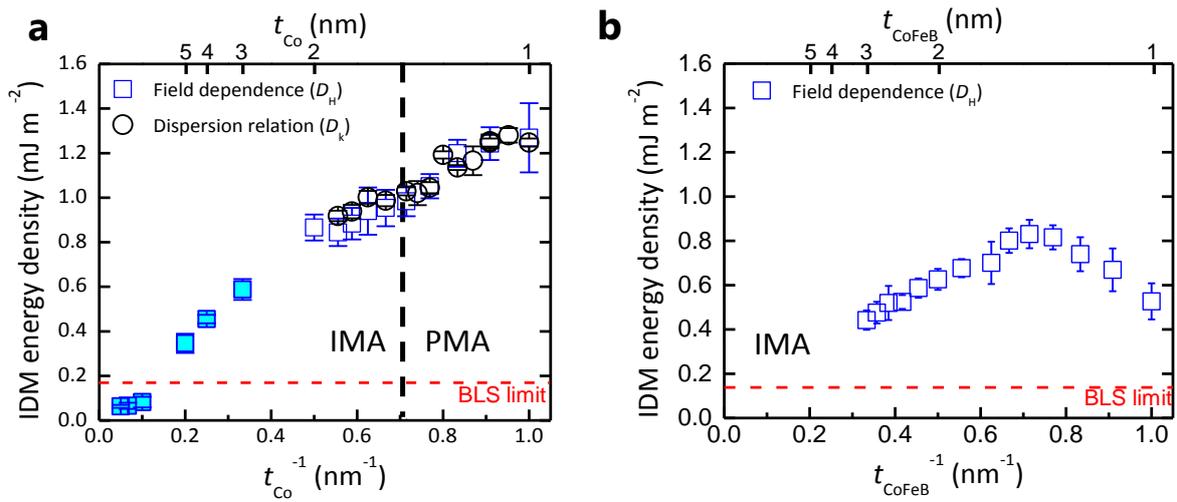

**Figure 5 | The IDM energy densities. a,** The IDM energy density as a function of $t_{Co}^{-1}$ for the two measurement methods. Blue squares and black circles show the IDM energy density measured by field dependence ($D_H$) and dispersion relation ($D_k$), respectively. The black dot line indicates the spin configuration changes PMA to inplane magnetic anisotropy (IMA). **b,** The IDM energy density as a function of $t_{CoFeB}^{-1}$ for the field dependence ($D_H$). The red dot lines show the limit for our BLS setup. Error bars correspond to s.d.

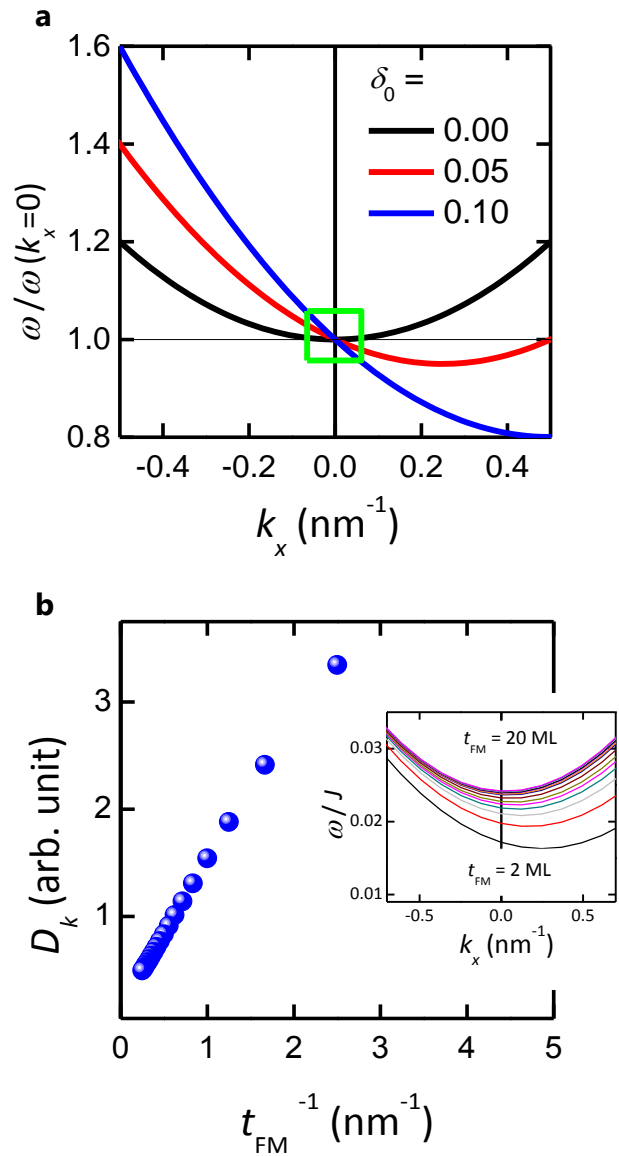

**Figure 6 | Numerically calculated SW dispersion relations. a,** Calculated SW dispersion relations for different IDM energy densities. The green box indicates that SW wavevector range of our BLS experiment. **b**, The IDM energy density from the SW dispersion relations as a function of $t_{FM}^{-1}$. Inset: Calculated SW dispersion relations for various $t_{FM}^{-1}$ from 2 ML to 20 ML when $\delta_0$ ($SD/J_{ex}a$) is 0.1. The vertical line indicate the $k_x = 0$. The SW frequencies at $k_x = 0$ increase with increasing $t_{FM}$.

# SUPPLEMENTARY FIGURES

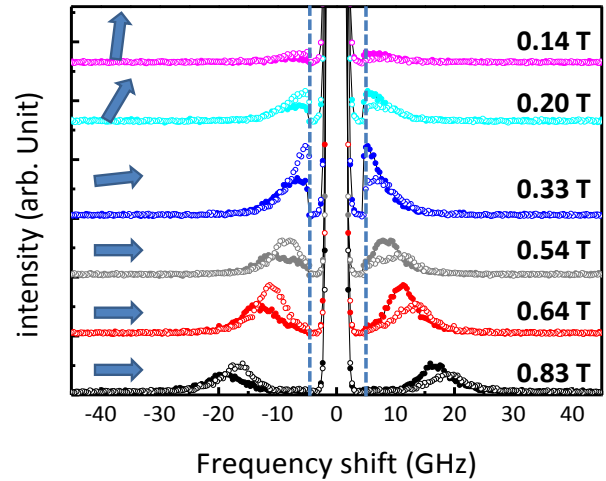

**Supplementary Figure 1 | SW spectra with their mirror images for small in-plane field (0.14 ~ 0.83 T).** When the in-plane magnetic field is larger than 0.5 T, the magnetization is out-of-plane, while it is smaller than 0.5 T, the magnetization is tilted. The schematic magnetization directions are shown as arrows in the left side of each spectrum. The dashed lines indicate Rayleigh scattering came from the interferometer shutter.

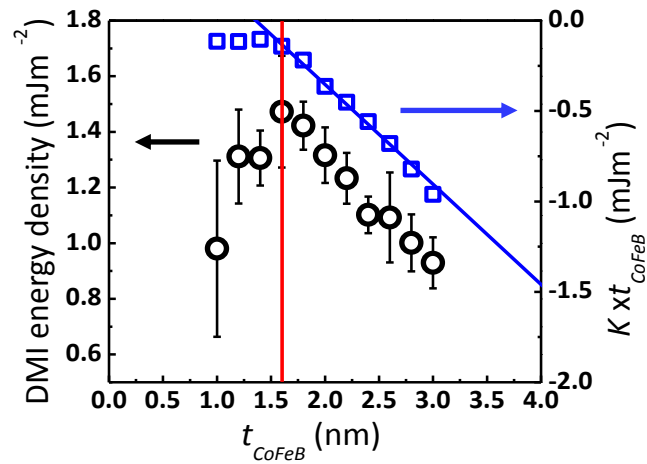

**Supplementary Figure 2 | DM energy density and $K_{eff} \times t_{CoFeB}$ as a function of $t_{CoFeB}$.**
When $t_{CoFeB} < 1.6$ nm, the $K_{eff} \times t_{CoFeB}$ starts to deviate from the linear behaviour, and DM interaction shows the same behaviour.

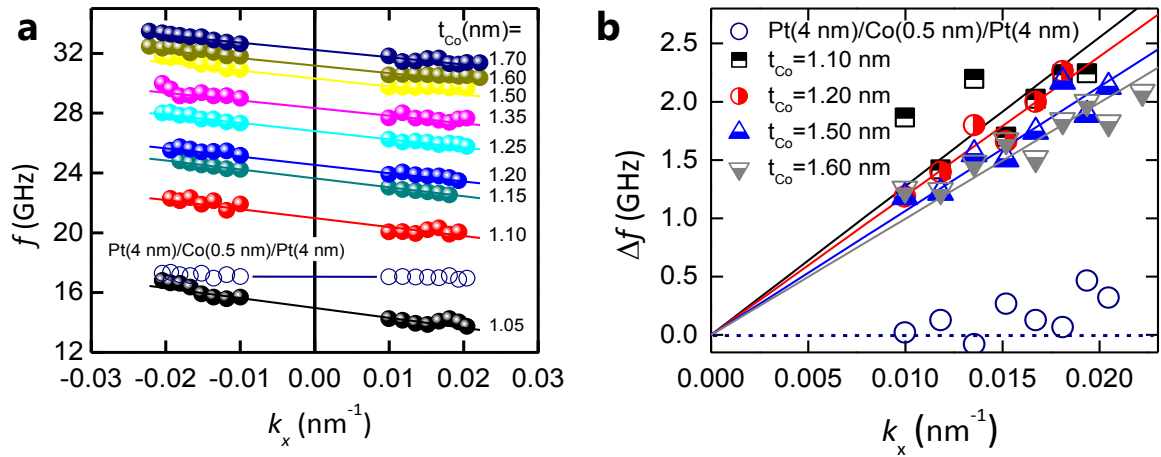

**Supplementary Figure 3 | The SW dispersion relation and the linearity of the frequency differences for different Co thicknesses. a**, The asymmetric dispersion relation measured by the BLS for various Co thicknesses. The open navy circles indicate the dispersion relation for a symmetric Pt (4 nm)/Co (0.5 nm)/Pt (4 nm) sample. For these measurements, the applied magnetic field is fixed at $H_{ext}$ = 0.915 T. **b**, All $\Delta f$ and linear fitting lines are visualized in one graph. The open navy circles show the $\Delta f$ for the symmetric Pt (4 nm)/Co (0.5 nm)/Pt (4 nm) sample.

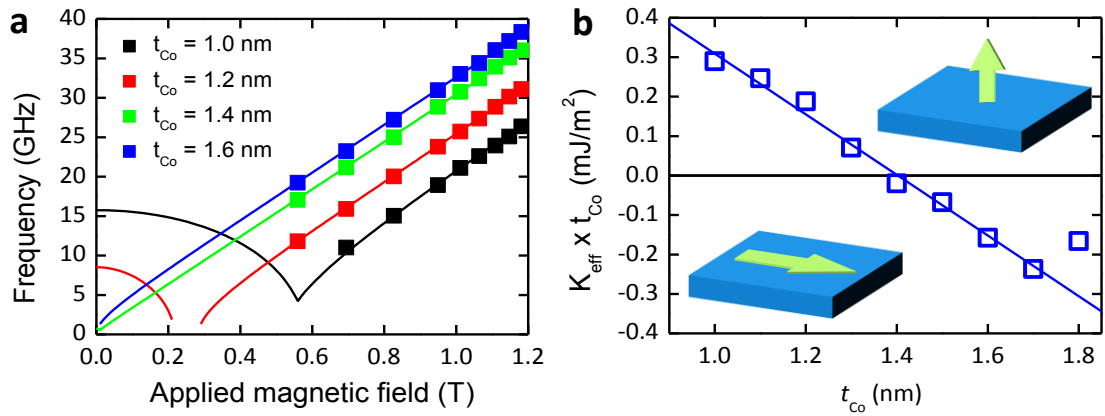

**Supplementary Figure 4 | Measured SW frequencies and PMA energy densities using BLS. a**, The field dependence SW resonance frequencies measured by BLS for various Co thickness. The squares indicate the median of the Stokes and anti-Stokes peak frequency and the solid lines are fits to Supplementary Eq. (3). **b**, $K_{eff} \times t_{Co}$ vs $t_{Co}$ plot with linear fit. We extract $K_s$ and $M_s$ from the slope and y-axis crossing. Above $t_{Co} > 1.4$ nm, the effective anisotropy becomes negative and the easy axis of the sample is in-plane.

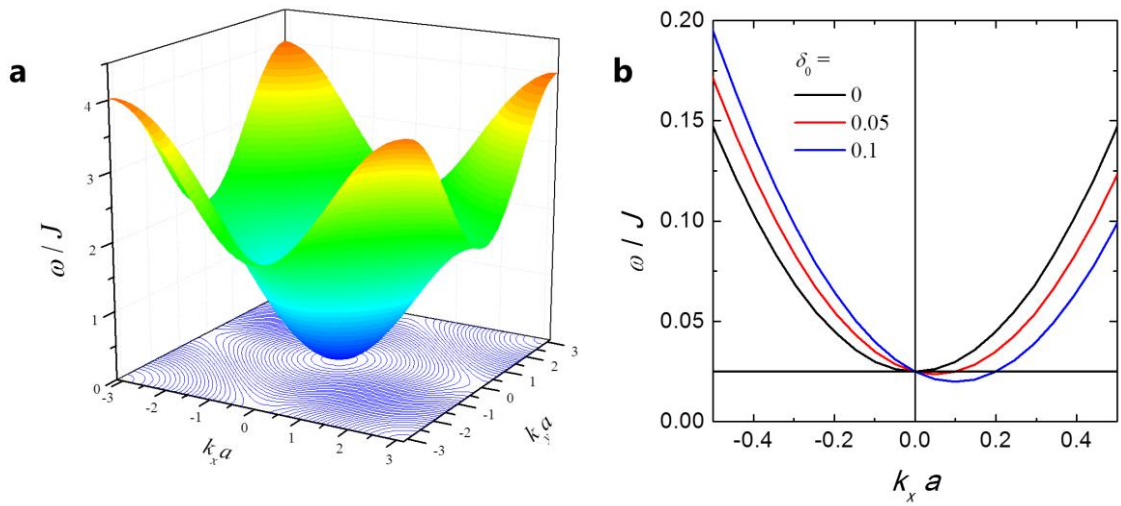

**Supplementary Figure 5 | Calculated SW dispersion relations including the IDM interaction. a**, Three dimensional SW dispersion relations for $k_x a$ and $k_y a$ without IDM interaction **b**, SW dispersion relations as a function of $k_x a$ with $k_y=0$ for $\delta_0 =0, 0.05, 0.1$.

# SUPPLEMENTARY NOTES

## Supplementary Note 1

**Brillouin light scattering (BLS) technique and interfacial Dzyaloshinskii-Moriya (IDM) interaction**

In order to determine the interfacial Dzyaloshinskii-Moriya (IDM) interaction, we measure the frequency difference ($\Delta f$) as a function of magnetic field or wavevector of the propagating spin wave (SW) by performing Brillouin light scattering (BLS).

All BLS data are governed by the so-called Damon-Eshbach (DE) mode including the contribution of the IDM interaction[1]:

$$f_{DE} = f_0\left(M_S, H_{ext}, K_U, A_{ex}, k_x\hat{\mathbf{x}}\right) + p\frac{\gamma D}{\pi M_S}k_x, \qquad (1)$$

where, $f_0$ is the SW frequency without the IDM contribution, $H_{ext}$, $K_u$, $A_{ex}$, $p$ and $k_x = \frac{4\pi}{\lambda}\sin\theta$ are the external magnetic field, the magnetic anisotropy, the exchange stiffness, the polarity of the magnetic field ($p = \pm 1$) and the wave vector of the incoming light where $\theta$ is the incident angle of the light, respectively.

For the case of the field dependence, only the $f_0$ term is varied in Supplementary Eq. (1), and $\Delta f$ does not vary. To check Supplementary Eq. (1) and reduce the uncertainty in the derived IDM energy density, we measure $\Delta f$ as a function of magnetic field. The observed constant $\Delta f$ as a function of the applied field validates the use of Supplementary Eq. (1) and allows us to accurately determine $\Delta f$, and hence, the derived IDM energy density.

During the measurements where we vary the $k$-vector, we fix the applied magnetic field at 0.915 T. As shown in Fig. 3b in the main text and as expected from Supplementary Eq. (1), $\Delta f$ varies linearly with the $k$-vector as described by:

$$\Delta f = \left|f_{DE}\left(+k_x\hat{\mathbf{x}}\right) - f_{DE}\left(-k_x\hat{\mathbf{x}}\right)\right| = \frac{2\gamma D}{\pi M_S}k_x, \qquad (2)$$

where $\gamma$ and $D$ are the gyromagnetic ratio and the IDM energy density, respectively. By linear fitting, we directly extract the IDM energy. We conclude that both magnetic field and $k$-vector dependent measurements are well described by Supplementary Eq. (1) and (2).

## Supplementary Note 2

**Advantages of BLS to determine the IDM interaction energy density**

Many different techniques are currently employed to study the IDM interaction, such as ferromagnetic resonance (FMR)[2], domain wall motion[3,4], and spin-polarized electron energy loss spectroscopy (SPEELS)[5]. However, FMR has not resulted in conclusive results as yet, and its ill-defined wavevector requires careful data analysis. Domain wall (DW) motion and nucleation based techniques are intrinsically complex as many aspects need to be taken into account e.g. DW energy profiles, pinning potential, etc. SPEELS can measure SW dispersion relations, however, the range of measurable wavevectors is 2 ~ 10 nm$^{-1}$ (corresponding to 0.1 ~ 0.5 nm length scale) is too limited to draw conclusions. Furthermore, SPEELS cannot measure the field dependence which is required to confirm the source of the $\Delta f$. Furthermore, it requires alternative means to measure the saturation magnetization required to extract the IDM energy density.

In contrast to the above methods, BLS uses a well-defined SW wavevector, which is determined by the wavelength and incident/scattering angle of the laser beam. Moreover, BLS is able to detect propagating SW excitations with $\pm k$-wavevectors simultaneously (Stokes and anti-Stokes peaks)[6,7,8]. The range of usable wavevectors is 0.01 ~ 0.02 nm$^{-1}$ corresponding to a length scale of 50 ~ 100 nm which is close to the length scale of skyrmions[9,10] with GHz range SW excitation. Utilizing the magnetic field and $k$-vector dependent BLS measurements the IDM energy density can be determined from the frequency difference between SWs with opposite ($\pm k$) wavevector. Furthermore, BLS allows for a direct measurement of the saturation magnetization (see Supplementary Note 6) and for local probing of samples as a small laser spot size is used. Specifically, we can perform local BLS measurements on ultrathin wedge shaped Pt/Co/AlO$_x$ and Pt/CoFeB/AlO$_x$ samples in order to investigate the thickness dependence of the IDM energy density as function of the Co (CoFeB) layer.

## Supplementary Note 3

**Asymmetric SW Dispersion for Out-Of-Plane Magnetization Geometry**

Asymmetric SW dispersion is a finger print of the IDM interaction when the magnetization is in-plane configuration. Our experimental conditions are satisfied this conditions. However, Cortés-Ortuño et al.[11] pointed out the asymmetry vanishes when the magnetization is out-of-plane. Therefore, it must be examined in our experiments by reducing the in-plane applied field. BLS SW spectra with various in-plane magnetic fields of the 1.2-nm-thick Co sample are shown in Supplementary Fig. 1. In this figure, the largest peak which occurs around 0 GHz is due to elastically scattered light, so-called "Rayleigh scattering", which is not related with magnetic signals. The peaks around 15 ~ 20 GHz with 0.83 T are typical BLS signals from the Pt/Co/AlO$_x$ sample. The closed circles are measured spectra and the open circles are mirror spectra in order to show clearly the frequency differences. We only show rather large fields (> 0.5 T) spectra in the manuscript due to the measurement limitation of our BLS system. The vertical dashed lines which indicate near Rayleigh scattering came from the interferometer shutter, are unavoidable. For the case of H$_{ext}$ < 0.33 T in our data, the spin orientations are changed from in-plane to out-of-plane (the blue arrows schematically indicates the magnetization directions). When the applied magnetic fields are less than 0.33 T, the peak position cannot be determined correctly, because of the shutter. Moreover, when the applied magnetic field is 0.14 T, the SW intensity is too small to confirm the correct peak positions. Therefore, unfortunately, we are not able to obtain meaningful spectra for fields smaller than 0.5 T, and this is the reason why we only show spectra for rather large fields where the magnetization direction is in-plane. Because of the limitation of our measurement system, we cannot determine whether the asymmetric dispersion is vanished for the out-of-plane magnetization or not.

## Supplementary Note 4

**The non-linear behaviour of frequency difference of Pt/CoFeB/AlO$_x$**

In Fig. 2b in the main text, the frequency difference of Pt/CoFeB/AlO$_x$ shows a maximum value at 1.6 nm, while Pt/Co/AlO$_x$ shows clear linear behaviour. Physical reason of such non-linear behavior of Pt/CoFeB/AlO$_x$ must be addressed. In order to resolve the un-expected behavior of Pt/CoFeB/AlO$_x$, we plot together $K_{eff} \times t_{CoFeB}$ and DM energy density via $t_{CoFeB}$ (thickness of CoFeB) in Supplementary Fig. 2. It is clear that the linear behaviour is broken in the $K_{eff} \times t_{CoFeB}$ vs. $t_{CoFeB}$ plot, when $t_{CoFeB} < 1.6$ nm. Based on our observation, we speculate the interface quality is changed due to the too thin ferromagnetic layer. Such deviation is usually observed in $K_{eff} \times t$ vs. $t$ plots for PMA materials (see Supplementary Note 6). The onset of the non-linear behaviour in the frequency difference or DM energy density is exactly the same thickness. Therefore, it implies that the non-linear behaviour in Fig. 2b in the main text is due to the degradation of the interface quality.

## Supplementary Note 5
**The SW *k*-vector dependent BLS measurements for a symmetric sample**.

In this section, we discuss the BLS measurements for nominal symmetric-interface samples such as Pt (4 nm)/Co (0.6 nm)/Pt (4 nm). As described in the main text, from this nominally symmetric structure we expect negligible or zero IDM interaction. SW *k*-vector dependent measurements are performed similarly as used for Fig. 4 of the main text and are shown in Supplementary Fig. 3. Open navy circles in Supplementary Fig. 3a indicate the SW dispersion relation. Due to the limited *k*-vector range, we only observe the symmetric dispersion, which implies a small IDM interaction. Supplementary Fig. 3b shows the correlation ($\Delta f = \left| f_{DE}(+k_x\hat{\mathbf{x}}) - f_{DE}(-k_x\hat{\mathbf{x}}) \right| = \frac{2\gamma D}{\pi M_S} k_x$) between the frequency differences and SW *k*-vector. No significant IDM interaction is observed by using BLS. To elucidate, two reasons are suggested; first, our examined system is more symmetric compared to the other reports (Refs. 8 and 13 in main text), and second, the IDM interaction might be small and cannot be detected by BLS as a small Δ*f* falls within the detection limit. Therefore, a small frequency, which indicates a small or negligible IDM energy density *cannot* be identified by BLS.

For the BLS measurements, a tandem interferometer with a free spectral range (FSR) of 75 GHz and a $2^8$ multichannel analyser is used. The frequency resolution in the measured Stokes and anti-Stokes peaks in the BLS spectra can be determined by using FSR/$2^8$ GHz. Therefore, the frequency resolution of the BLS setup is approximately 0.29 GHz. Since the correlation between the frequency difference and the IDM energy density is given by $\Delta f = \frac{2\gamma D}{\pi M_S} k_x$, we can simply deduce that the resolution of the obtained IDM energy density is about $D = 0.164$ mJm$^{-2}$ with a saturation magnetization of $M_s = 1100$ kAm$^{-1}$, $\gamma = 2.37 \times 10^{11}$ T$^{-1}$s$^{-1}$, $k_x = 0.0167$ nm$^{-1}$, and Δ*f* = 0.29 GHz, respectively.

## Supplementary Note 6
**Determination of the saturation magnetization and anisotropy energies**

In this section, we demonstrate the SW dispersion relation without the IDM interaction. First, in order to define the SW frequency without the IDM interaction, the median value of the Stokes and anti-Stokes peak are taken to determine the perpendicular magnetic anisotropy energy and the saturation magnetization. As illustrated in Supplementary Fig. 4a, the applied magnetic field dependence of SW are measured by BLS for various Co thicknesses ($t_{Co}$ = 1.0, 1.2, 1.4, and 1.6 nm). Since the applied magnetic field is perpendicular to the magnetization creating the surface SW mode, the SW excitation frequencies are given by[12]:

$$f_0 = \frac{\gamma}{2\pi}\sqrt{\left[H_{ext}\cos\theta - \left(4\pi M_S - \frac{2K_u}{M_S}\right)\sin^2\theta\right]\left[H_{ext}\cos\theta - \left(4\pi M_S - \frac{2K_u}{M_S}\right)(\cos^2\theta - \sin^2\theta)\right]} \quad (3)$$

where, $\gamma$ is the gyromagnetic ratio (= $2.37\times10^{11}$ T$^{-1}$s$^{-1}$), $\theta$ is the angle between the magnetization and the sample plane, $K_u$ is the perpendicular uniaxial anisotropy constant, $H_{ext}$ is the external magnetic field, $M_s$ is the saturation magnetization, respectively. In this equation, the contributions of dipolar field and exchange energy have been neglected as is justified in the ultrathin limit. Consequently, the measured SW frequencies and the fitted curves show a good correspondence as shown in Supplementary Fig. 4a. For the case of $t_{Co} >$ 1.4 nm, the frequencies of the propagating SWs differ from the thinner thicknesses, which means that the effective uniaxial anisotropy ($K_{eff} = 2K_s/t - 1/2\mu_0 M_s^2$) is changed from positive (out-of-plane) to negative (in-plane) values. To elaborate, we plot the anisotropy energy density ($K_{eff} \times t_{Co}$) as a function of $t_{Co}$ in Supplementary Fig. 4b. From this plot, we determine the slope and y-crossing, corresponding to the volume anisotropy ($-1/2\mu_0 M_s^2$) and the surface anisotropy ($K_s$)[13], respectively. This allows us to extract $M_s$ directly from the BLS measurements. $M_s$ is the only necessary physical quantity to convert the measured $\Delta f$ to the IDM energy density. The obtained $K_s$ is 0.54 mJm$^{-2}$ and $M_s$ is 1100 kAm$^{-1}$, which is about 78.5% of the bulk Co value.

## Supplementary Note 7
**Double time Green's function for SW dispersion relations**

To calculate the SW dispersion relation in ultrathin ferromagnetic layers and super-lattices, the double time Green's function method is widely used[14,15,16,17,18]. It is a well-developed method in statistical physics[19] and magnetism[20].

We now briefly review the double time Green's function to obtain the SW dispersion relations of N atomic ferromagnetic layers with DM interaction. The Hamiltonian with DM interaction in terms of the spin operator is given by[21, 22]:

$$H = -g\mu_B H_{ext} \sum_i S_i^z - \sum_{\langle i,j \rangle} J_{ij} \mathbf{S_i} \cdot \mathbf{S_j} - K_u \sum_{\langle i,j \rangle} S_i^z S_j^z - K_s \sum_{\langle i,j \rangle'} S_i^z S_j^z + \sum_{\langle i,j \rangle''} D_{ij} \hat{\mathbf{z}} \cdot (\mathbf{S_i} \times \mathbf{S_j}), \quad (4)$$

where, $J_{ij}$ and $D_{ij}$ are the isotropic inter-atomic Heisenberg and DM exchange energies between the $i$-th and $j$-th spins, and $K_u$ and $K_s$ are the bulk and surface uniaxial anisotropy energies. $\langle i,j \rangle, \langle i,j \rangle', \langle i,j \rangle''$ denote the summations of the nearest neighbours. The last DM interaction term can be rewritten as

$$H_{DMI} = \sum_{\langle i,j \rangle''} D_{ij} \left( S_i^x S_j^y - S_i^y S_j^x \right). \quad (5)$$

Following Ref. 16, the double time Green function can be defined as:

$$G_{ij}(t,t') = \langle\langle b_i^+(t) | b_j^-(t') \rangle\rangle. \quad (6)$$

The equation of motion for $G_{ij}$ is

$$i \frac{dG_{ij}(t,t')}{dt} = \langle [b_i^+(t), b_j^-(t')] \rangle \delta(t-t') - \langle\langle [H, b_i^+(t)] | b_j^-(t') \rangle\rangle, \quad (7)$$

and the higher order Green's functions are decoupled by the random phase approximations, the set of differential equations for $N$-atomic ferromagnetic layers can be obtained[22].
We define the normalized energy quantities as $\delta_0 = D/J$, $k_u = K_u/J$, $k_{s1} = K_{s1}/J$, and $k_{sN} = K_N/J$. $k_x$, $k_y$, and $a$ are the $x$ and $y$ component of the SW vector and the lattice constant, respectively. In these calculations, we assume a simple cubic lattice structure, but this model can be extended for $bcc$ and $fcc$ structures[22]. The DM interaction contribution is developed with the number operator, $\hat{n}_i = b_i^+ b_i^-$ [16]:

$$\sum_{i,j} D_{ij}\left(S_i^x S_j^y - S_i^y S_j^x\right) = \frac{1}{4}\sum_{i,j} 2iD_{ij}\left(b_i^- b_j^+ - b_i^+ b_j^-\right) \tag{8}$$

$$\cong D\sum_i \left(\sin(k_x a) + \sin(k_y a)\right)\hat{n}_i. \tag{9}$$

From the matrix equations, we obtain the series of the Green's functions and eigenvalues of $\Delta(E)$ for a given *k*-vector. The *N* eigenvalues correspond to SW excitation energies, and corresponds with the SW dispersion relation. The typical SW excitations for the lowest energies are shown in Supplementary Fig. 5a and b for $N = 2$ simple cubic ferromagnetic layers with $k_u = 0.01$, $k_{s1} = k_s N = 0.01$ and $\delta_0 = 0, 0.05, 0.1$.